\documentclass[12pt,preprint]{aastex}

\lefthead{Bamba et al.}
\righthead{Diffuse Hard X-ray sources with {\it ASCA}}

\begin{document}

\title{
Diffuse Hard X-ray Sources Discovered  with the {\it ASCA} Galactic Plane Survey}

\author{
Aya Bamba, Masaru Ueno, and Katsuji Koyama,
}
\affil{
Department of Physics, Graduate School of Science, Kyoto University, 
Sakyo-ku, Kyoto 606-8502, Japan
}
\email{bamba@cr.scphys.kyoto-u.ac.jp,
masaru@cr.scphys.kyoto-u.ac.jp,
koyama@cr.scphys.kyoto-u.ac.jp}

\and

\author{Shigeo Yamauchi}
\affil{
Faculty of Humanities and Social Sciences, Iwate University, 
3-18-34 Ueda, Morioka, Iwate 020-8550, Japan
}
\email{yamauchi@iwate-u.ac.jp}

\begin{abstract}

We found diffuse hard X-ray sources, G11.0+0.0, G25.5+0.0, and 
G26.6$-$0.1 in the {\it ASCA} Galactic plane survey data.
The X-ray spectra are featureless with no emission line,
and are fitted with both models of
a thin thermal plasma in non-equilibrium ionization
and a power-law function.
The source distances are estimated to be 1--8~kpc,
using the best-fit $N_{\rm H}$ values on the assumption
that the mean density in the line of sight is 1~H~cm$^{-3}$.
The source sizes and luminosities are then 4.5--27~pc and 
(0.8--23)$\times10^{33}$~ergs~s$^{-1}$.
Although the source sizes are typical to supernova remnants (SNR)
with young to intermediate ages,
the X-ray luminosity, plasma temperature, and
weak emission lines in the spectra are all unusual.
This suggests that these objects are either shell-like SNRs
dominated by X-ray synchrotron emission, like SN~1006, or, alternatively, 
plerionic SNRs.
The total number of these classes of SNRs in our Galaxy is also estimated. 

\end{abstract}


\section{Introduction}

Since the discovery of cosmic rays \citep{hess},
the origin and the acceleration mechanism up to TeV or even more
have been long-standing problems.
The plausible acceleration sites in our Galaxy are
shell-type SNRs with diffusive shock acceleration mechanism
\citep{wentzel,reynolds1998}, pulsars and pulsar wind nebulae 
\citep{dejager},
although \citet{reynolds1999} suggests that 
no SNR can accelerate electrons up to $10^{15.5}$~eV
(the ``knee'' energy).
Both of them are characterized by hard X-rays via synchrotron radiation
and we have already several samples of such SNRs:
SN~1006 and some other SNRs which accelerate cosmic rays on their shell
\citep{koyama1995,koyama1997,slane1999,bamba2000,borkowski2001b,slane2001}
and the plerionic SNRs
\citep{asaoka,weisskopf,blanton}.
Together with the discoveries of inverse Compton $\gamma$-rays,
which are up-scattered cosmic microwave photons or synchrotron X-rays
by the high energy electrons
\citep{tanimori,muraishi,enomoto,weekes},
the hard X-ray emission supports
the presence of extremely high energy electrons.
The total and maximum energy of accelerated electrons in
SN~1006 has been studied  by {\it ASCA} \citep{allen,dyer} 
and {\it Chandra} \citep{bamba2003}.
In order to account for the total flux of the Galactic high energy 
cosmic rays, a large numbers of SNRs with non-thermal X-rays
must be still hidden in the Galactic plane
and should be discovered.

The maximum electron energy in the SNRs is determined
by the balance between the acceleration and radiative energy loss.
The acceleration rate is proportional to the magnetic field strength ($B$). 
However, the synchrotron energy loss is proportional to $B^2$ and
electron energy.
Ages of SNRs, densities of circumstellar matter,
and explosion energies also limit the maximum energy.
Since the synchrotron X-rays are produced by electrons with higher energy 
than those responsible to the radio emission,
SN~1006-like SNRs should have relatively weak magnetic field ($B$),
and they must be generally radio faint SNRs like SN~1006.
In fact, new SN~1006-like SNRs such as G347.3$-$0.5, G266.2$-$1.2,
and G28.6$-$0.1 are first found with X-rays,
then later identified as radio faint SNRs
\citep{koyama1997,slane1999,bamba2001a,koyama2001}.
For the plerionic SNRs,
the relation between the luminosity in X-ray and radio band is
an unknown problem
\citep{hands}.
For example, 
3C58 has the radio luminosity 1000 times smaller than the Crab Nebula
although both are strong X-ray emitters
\citep{green}.
Accordingly, the previous search for SNRs based on
the radio observations \citep{green} may not be the most optimized method.

We thus have searched for diffuse hard X-ray sources
in the Galactic plane survey with {\it ASCA} \citep{sugizaki1999},
and found a handful of SNR candidates with hard X-rays.
We have carried out long exposure observations on the selected 4 candidates
to investigate whether or not these are new SNRs.
One of the candidates, AX~J1843.8$-$0352,
is already identified as a new SN~1006-like SNR \citep{bamba2001a}
with the aid of the radio continuum emission of G28.6$-$0.1 \citep{helfand}
and the result is confirmed by
the {\it Chandra} follow-up observations \citep{koyama2001,ueno}.

This paper reports on the results on the other three candidates,
(G11.0+0.0, G25.5+0.0, and G26.6$-$0.1) in \S~\ref{analyses},
and discusses the mechanisms of their X-ray emissions
(see \S~\ref{discuss1} and \S~\ref{discuss2}).
The implications on the total number of SNRs with hard X-rays in our Galaxy
are also discussed in \S~\ref{discuss3}.

\section{Observations and Data Reduction}

The {\it ASCA} Galactic plane survey has covered the area of the Galactic plane of
$|l|\leq 45\fdg0$ and $|b|\leq 0\fdg4$,
with the exposure time of $\sim 10$~ks each.
In the mosaic map shown by \citet{sugizaki1999},
we searched for SNRs in hard X-rays
excluding about 10\% of the surveyed area,
which was suffered from stray lights of bright X-ray sources.
We selected the 4 brightest diffuse hard X-ray sources 
and performed follow-up deep exposure observations.
The observation dates and the on-axis positions are summarized
in Table~\ref{obslog}.

{\it ASCA} carried four XRTs (X-Ray Telescopes; Serlemitsos et al.\ 1995)
with two GISs (Gas Imaging Spectrometers; Ohashi et al.\ 1996) and 
two SISs (Solid-state Imaging Spectrometers; Burke et al.\ 1994)
on the focal planes.
Since our targets are diffuse sources
with sizes comparable to the SIS field of view (FOV),
we do not refer to the SIS data in this paper.
In all of the observations, GISs were operated in the nominal PH mode.
We rejected the GIS data obtained in the South Atlantic Anomaly,
in low cut-off rigidity regions ($<6$~GV), 
or when the target's elevation angle was low ($< 5^\circ$).
Particle events were removed by the rise-time discrimination method
\citep{ohashi}.
After these screenings, the total available exposure time of each observation
is shown in Table~\ref{obslog}.
To increase the statistics, the data of the two detectors, GIS-2 and GIS-3,
were co-added in the following study.

\section{Analyses and Results}
\label{analyses}

\subsection{X-Ray Images and Source Detection}

The  GIS images in the 0.7--7.0~keV band around the SNR candidates
are shown in Figure~\ref{images}.
Diffuse enhancements can be seen with the centers at
about ($l$, $b$) = (11\fdg0, 0\fdg0), (25\fdg5, 0\fdg0),
and (26\fdg6, $-$0\fdg1), hence are referred to as G11.0+0.0, G25.5+0.0, 
and G26.6$-$0.1.

Weak point-like sources No.1, 7, and 12 are
detected in G11.0+0.0, G25.5+0.0, and G26.6$-$0.1, respectively.
We also checked the {\it ROSAT}%
\footnote{Available at http://www.xray.mpe.mpg.de/cgi-bin/rosat/src-browser.}
and $Einstein$%
\footnote{Available at http://heasac.gsfc.nasa.gov/docs/einstein/archive/heao2\_archive.html.}
point source catalogs and found only one source,
1RXS~J184049.1$-$054336 in G26.6$-$0.1 (the same source as No.12)
from 15~arcmin circles of the diffuse sources.
The flux of these point-like sources contributes
about 10\% of the diffuse flux (see Appendix).
We thus made the radial profiles in the 0.7--7.0~keV band 
including the point-like sources,
then fitted with a Gaussian plus constant (for the background) model
(Figure~\ref{profiles}).
The diffuse components are detected
with the confidence level of 10.2$\sigma$ for G11.0+0.0,
5.7$\sigma$ for G25.5+0.0, and 7.5$\sigma$ for G26.6$-$0.1, respectively.
The best-fit FWHM is 12.2 ($>$6.3), 10.8 ($>$4.9),
and 8.9 ($>$5.7)~arcmin for G11.0+0.0, G25.5+0.0, 
and G26.6$-$0.1, respectively
(parentheses indicate single-parameter 90\% confidence regions).
The FWHM sizes are significantly larger than
that of the point spread function of {\it ASCA} GIS (3~arcmin),
hence we confirm the diffuse nature
for G11.0+0.0, G25.5+0.0, and G26.6$-$0.1.

To confirm the {\it ASCA} results, 
we checked the {\it ROSAT} all sky survey images%
\footnote{Available at http://www.xray.mpe.mpg.de/cgi-bin/rosat/rosat-survey.}.
However, no counterparts of G11.0+0.0 nor G25.5+0.0 is found,
mainly because of the large absorption and/or short exposure time.
As for G26.6$-$0.1, a {\it ROSAT} counterpart is found. However, 
no pointing observation was made.
Thus all the available data with reasonable spatial resolution and exposure
are provided solely from our {\it ASCA} observations,
hence we concentrate on the {\it ASCA} results hearinafter.

\subsection {X-ray Spectra}

For the spectral analyses,
we combined the survey and the follow-up observation data: 
obs.1 and 2 for G11.0+0.0, obs.3 and 4 for G25.5+0.0, and
obs.5 and 6 for G26.6$-$0.1 (see Table~\ref{obslog}).
The spectra were made using the photons in the circular regions
of diameters 15~arcmin for G11.0+0.0,
and 12~arcmin for both G25.5+0.0 and G26.6$-$0.1, respectively,
which are equal to $\sim 3\sigma$ width of the radial profiles
(Figure~\ref{profiles}).
These source regions are shown with the solid circles in Figure~\ref{images}.
X-ray photons of point sources, No.1 for G11.0+0.0, No.7 for G25.5+0.0, and 
No.12 for G26.6$-$0.1, were extracted from a 3~arcmin radius circle
each and were removed from the diffuse source data.

In order to properly subtract the Galactic ridge X-rays
\citep{koyama1986,kaneda},
the background data were extracted from the source free regions
near to the targets with the same FOVs and at the same distance from
the Galactic plane (dashed lines in Figure~\ref{images}).
For the background of G11.0+0.0, we selected the annual region
between 9 and 12 arcmin radius,
but excluded the circular regions of
6~arcmin radius around the sources No.3 and 4.

\begin{figure}[hbtp]
\epsscale{0.3}
\plotone{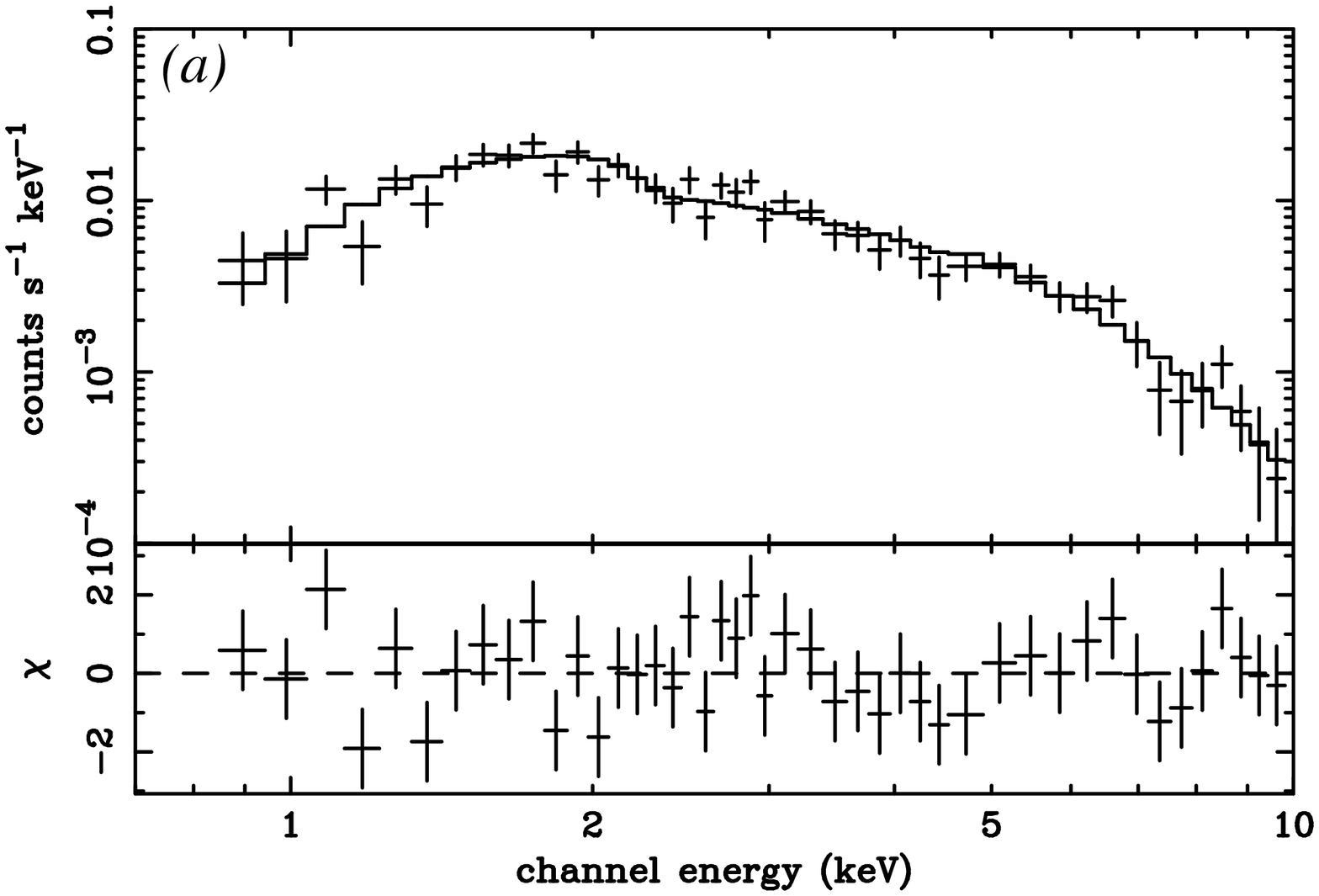}
\plotone{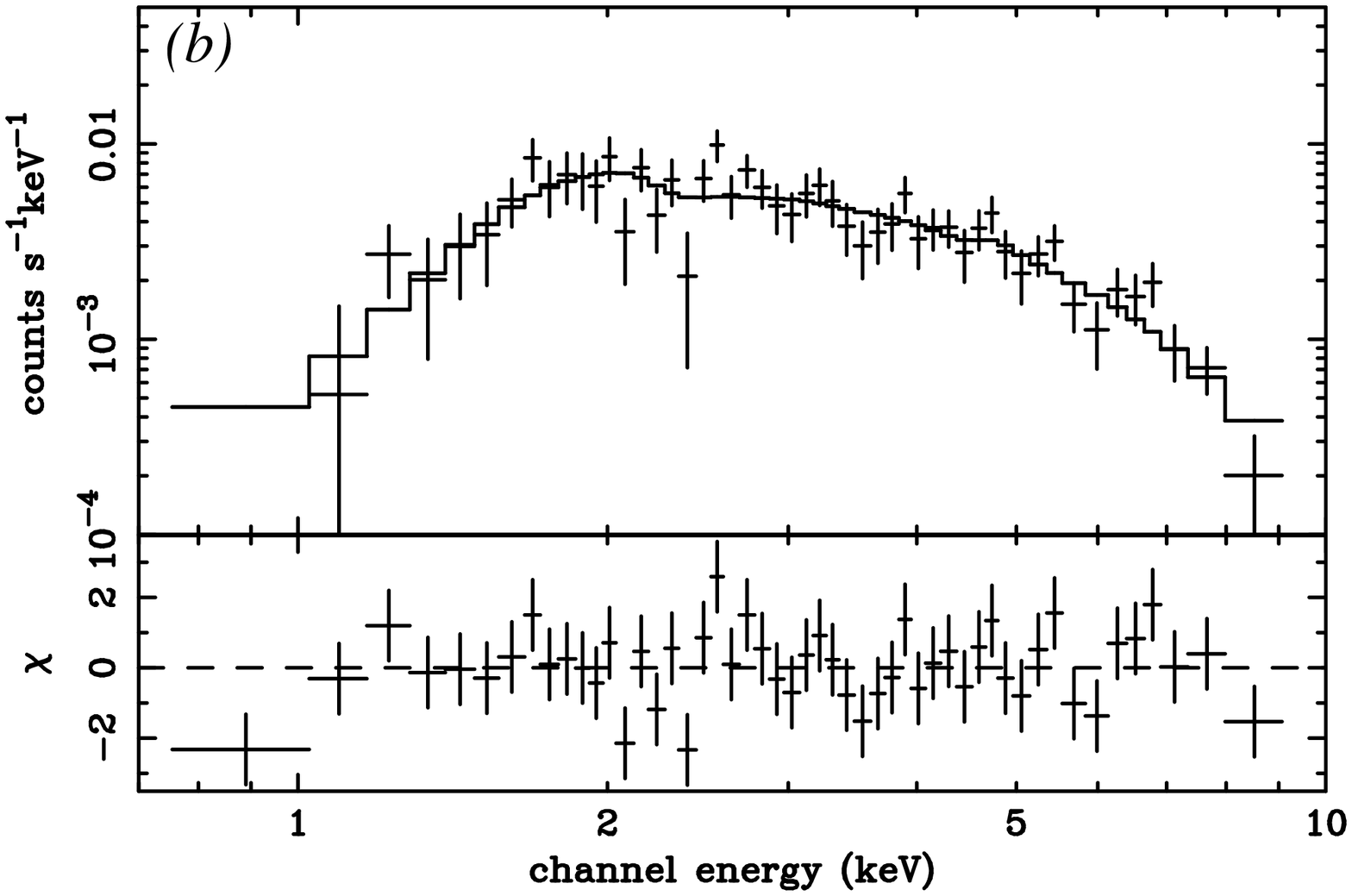}
\plotone{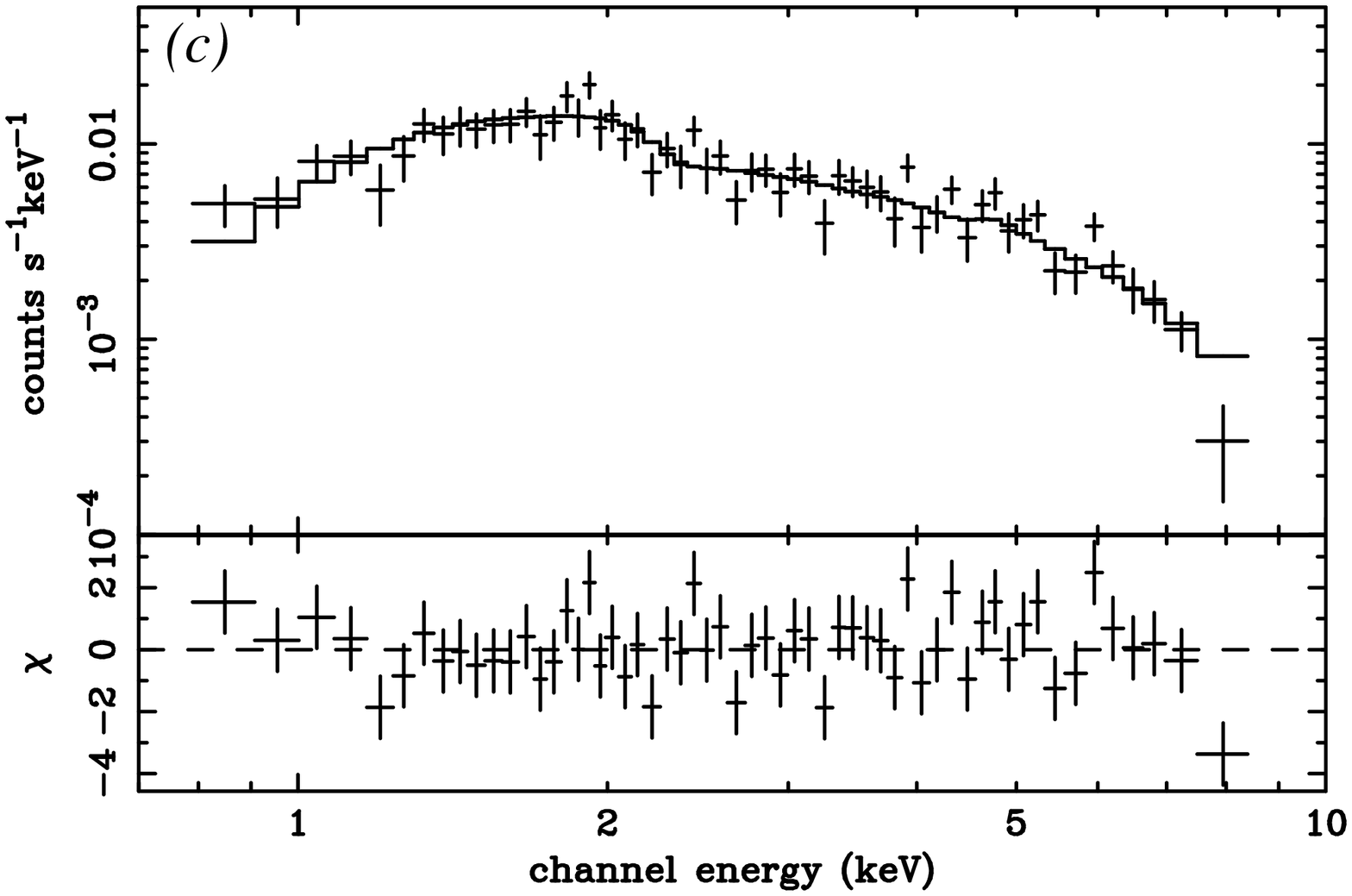}
\caption{Upper panels: Background-subtracted spectra (crosses).
(a: G11.0+0.0; b: G25.5+0.0; c: G26.6$-$0.1). 
The spectra  of  GIS~2 and 3 are combined. The solid lines
are  the best-fit power-law models.
Lower panels: The data residuals from the best-fit models.
\label{spectra}}
\end{figure}

The background-subtracted spectra are shown in Figure~\ref{spectra}.
We see relatively flat spectra and no line feature.
The spectra were fitted with two models of a thin thermal plasmas 
in non-equilibrium ionization (NEI) \citep{borkowski2001a},
and a power-law function.
The absorption column densities are calculated using the cross sections by
\citet{morrison} of solar abundances \citep{anders}.
These two models are statistically acceptable for all the spectra.
The best-fit parameters are given in Table~\ref{bestfit_para}, 
while the best-fit power-law models are shown in Figure~\ref{spectra} 
with the solid-line histograms.

\subsection {Physical Parameters of the Diffuse Sources}

We searched for radio, optical, and infrared (IR) counterparts
from SIMBAD data base at the positions of the diffuse X-ray sources,
G11.0+0.0, G25.5+0.0, and G26.6$-$0.1, but found no candidates.
We therefore can not derive the source distances
by the aid of the other wavelength information,
and have to rely solely on our  X-ray spectra.
The absorption column densities of G11.0+0.0 and G26.6$-$0.1
are significantly smaller
than, while that of G25.5+0.0 is comparable to that through
the Galactic inner disk ($\sim 10^{23}$~H~cm$^{-2}$; Ebisawa et al.\ 2001).
Therefore, the former two sources
are likely located at the near side of the Galactic plane,
while the latter would be in the Galactic Scutum arm.

To be more quantitative, we assumed that the average density
in our Galaxy toward the inner disk is 1~H~cm$^{-3}$ \citep{ebisawa}
and estimated the distances using the best-fit absorption column densities
in Table~\ref{bestfit_para}.
The results together with the source diameters and total luminosities are
listed in Table~\ref{phys_para}.
Although the results are based on the best-fit power-law model,
essentially no difference is found from those based on the NEI model. 

\section{Discussions}

From the Table~\ref{phys_para},
we conclude that G11.0+0.0, G25.5+0.0, and G26.6$-$0.1 are 
diffuse Galactic sources with large sizes of 5--30 pc
and moderate luminosities of $10^{33-34}$ erg s$^{-1}$.
For the nature of these diffuse sources,
a key question is whether the spectra are thermal or not.
Unfortunately, the observed spectra can not constrain on this issue
in the statistical point of view.
We hence discuss both the cases in the following subsections. 

\subsection{Are Diffuse Sources Thermal ?}
\label{discuss1}

If the spectra are thermal, 
one possibility for the diffuse sources is unresolved emissions 
of many young stars in star forming regions.
The typical example of this class, 
the Orion nebula, has the total X-ray luminosity of $10^{33}$~ergs~s$^{-1}$
and mean temperature of 3--4~keV \citep{yamauchi},
which are in the error ranges of the diffuse sources
(Table~\ref{bestfit_para} and Table~\ref{phys_para}).
Since G11.0+0.0 and G26.6$-$0.1 are in the near side of the Galactic ridge,
they should be detected as emission nebulae or H~$_{\rm II}$ regions,
but have no optical, IR, nor radio band counterparts.
On the other hand, since G25.5+0.0 is at a far distance
in the Galactic inner arm,
no detection in the other wavelength may not be surprising.
However the source size of $\sim$30~pc is exceptionally large
as a star forming region.
We thus conclude that the diffuse sources are not
likely to be star forming regions.

Another possibility of the nature of G11.0+0.0, G25.5+0.0, and G26.6$-$0.1 are 
either shell-like SNRs in the adiabatic phase or mixed morphology
SNRs of young to middle age (10$^{2-3}$ years). 
Usual SNRs of this class emit thermal X-rays with the temperature $<$~1~keV.
However, the spectra of G11.0+0.0 and G26.6$-$0.1 are very hard
($\sim$ several keV)
and have no strong emission line from highly ionized ions,
which are not in favor of thermal SNRs.
On the other hand, G25.5+0.0 has
relatively mild temperature and reasonable abundance.

Assuming a uniform density plasma sphere,
we estimate the emission measures $(E.M.)$,
dynamical times ($t$=radius/sound velocity),
electron densities $(n_{\rm e})$, total masses $(M)$,
and thermal energies $(E)$,
which are listed in Table~\ref{thermal}.
We find that total masses of G11.0+0.0 and G26.6$-$0.1 are
only a few solar, consistent with young SNRs of high temperature plasma.
However the total energies are significantly smaller than
the canonical supernova explosion.
Therefore only small fraction of the explosion energy must be  
converted to the thermal energy, hence they may be in the earlier phase
than adiabatic.
In this case, the X-ray emitting plasma 
should be attributable to the SN ejecta
with large metalicity,
but the observations show no large metal excess.
Thus thermal SNRs are unlikely for these diffuse sources.

For G25.5+0.0, the physical parameters are consistent with
the adiabatic phase SNR, however the plasma density ($<$ 0.1~H~cm$^{-3}$) is 
too small as the interstellar gas in the Scutum Arm.
We thus regard that a thermal SNR for this diffuse source
is also unlikely.     

\subsection{Non-Thermal Diffuse Sources ?}
\label{discuss2}
  
We then discuss a possibility of non-thermal origin.
The most likely case is SNRs with non-thermal X-rays,
like SN~1006 and the Crab Nebula.
In Table~\ref{phys_para}, we also list the physical parameters of
a newly established SN~1006-like SNR,
AX~J1843.8$-$0352 = G28.6$-$0.1 \citep{bamba2001a},
for comparison.

First, we discuss the possibility that
they are SN~1006-like SNRs.
Photon indices of G11.0+0.0 and G25.5+0.0 are $\sim 2$,
which are smaller than those of typical SN~1006-like SNRs, 
SN~1006 itself, G347.3$-$0.5, and G266.6$-$1.2,
but are similar to that of G28.6$-$0.1.
This may indicate that the synchrotron X-rays are due
to the electrons with higher energy than that of the acceleration cut-off,
which is expected from the first-order Fermi acceleration 
(expected $\Gamma$ = 1.5; Wentzel 1974).
The diameters and total luminosities also resemble
those of G28.6$-$0.1.
These facts strongly support that
at least G11.0+0.0 and G25.5+0.0 are SN~1006-like SNRs.

To establish the SN~1006-like SNRs,
the presence of synchrotron radio emission is essential.
However no cataloged radio source is found in
the NRAO VLA Sky Survey 20~cm survey archival data%
\footnote{Available at http://www.cv.nrao.edu/.}.
If the magnetic field is weaker than the other SN~1006-like SNRs,
the radio surface brightness would be below the current 
detection limit.

In order to estimate the radio flux band for  
the SN~1006-like SNRs, we tried the spectral fittings with 
a $srcut$ model \citep{reynolds1998,reynolds1999}.
The spectral index at 1~GHz was frozen to $\alpha = -0.5$,
expected value by first-order Fermi acceleration.
The fittings were statistically accepted 
and the best-fit parameters are listed in Table~\ref{bestfit_para}.
We also tried the same fittings with $\alpha = -0.6$
and found no essential difference in the best-fit parameters.
The best-fit flux density for each source suggests that
the expected surface brightness at 1~GHz is
9.8$\times 10^{-24}$,
1.7$\times 10^{-23}$,
and 9.8$\times 10^{-24}$~W~m$^{-2}$Hz$^{-1}$sr$^{-1}$
for G11.0+0.0, G25.5+0.0, and G26.6$-$0.1, respectively.
On the other hand,
the minimum surface brightness of the cataloged radio SNRs
in our survey field is  $2\times 10^{-21}$~W~m$^{-2}$Hz$^{-1}$sr$^{-1}$
(for G3.8+0.3; Case \& Bhattacharya 1998),
which is similar to  SN~1006
(3$\times 10^{-21}$~W~m$^{-2}$Hz$^{-1}$sr$^{-1}$;
Long, Blair, \& van den Bergh 1988; Winkler \& Long 1997)
and G347.3$-$0.5
($4\times 10^{-21}$W~m$^{-2}$Hz$^{-1}$sr$^{-1}$;
Ellison, Slane, \& Gaensler 2001; assuming $\alpha$ = $-0.5$), but
is two or three orders of magnitude larger than our new sources.
Even the minimum surface brightness of the all cataloged radio SNRs is
larger than our sample,
$6.2\times 10^{-23}$~W~m$^{-2}$Hz$^{-1}$sr$^{-1}$
(for G156.2+5.7; Case \& Bhattacharya 1998).
Therefore, no radio counterpart for our new X-ray sources would be 
simple due to limited detection threshold of the current radio observations.

The power-law spectra
and the luminosities ($10^{34}$--$10^{35}$~ergs~s$^{-1}$)
suggest that the new sources are also Crab-like (plerionic) SNRs,
although no radio pulsar is found \citep{chevalier}.
The luminosity ratio between the  radio and X-ray band
is largely different among the plerionic SNRs:
the radio luminosity of 3C58, for example is
about 10 times smaller, while that of X-ray is about 1000 times 
smaller than that of the Crab Nebula \citep{hands}.
We estimated the expected surface brightness
using the flux density of 3C58 and the Crab Nebula \citep{green}.
For the 3C58-like case,
the surface brightness of G11.0+0.0, G25.5+0.0, and G26.6$-$0.1 
at 1~GHz is expected  to be
2.5$\times 10^{-20}$,
5.1$\times 10^{-20}$,
and 1.9$\times 10^{-20}$~W~m$^{-2}$Hz$^{-1}$sr$^{-1}$, respectively.
Since these values are larger than the detection limit in the surveyed region
(see previous paragraph), we exclude the 3C58-like case.
On the other hand, for the Crab-like case, 
their respective surface brightness becomes to be
7.3$\times 10^{-23}$,
1.5$\times 10^{-22}$,
and 5.6$\times 10^{-23}$~W~m$^{-2}$Hz$^{-1}$sr$^{-1}$.
Therefore, a possibility of the Crab-like case can not be excluded.

The photon index of G26.6$-$0.1 is unreasonably small as a usual SNR.
The diameter and the total luminosity are also smaller than
those of G28.6$-$0.1 and the other SNR candidates with non-thermal X-rays.
Thus G26.6$-$0.1 may be of different class.
Similar nature to G26.6$-$0.1 is found with hard X-ray clumps \citep{uchiyama}
of $\gamma$ Cygni,
an SNR interacting with molecular clouds \citep{fukui}
and an EGRET source \citep{esposito};
all the sizes ($\sim$ a few pc), photon indices ($\sim$ 1.2),
and luminosities
($\sim 4\times 10^{32}$~ergs s$^{-1}$ in the 2.0--10.0~keV band)
of the clumps resemble those of G26.6$-$0.1.
\citet{uchiyama} concluded the mechanism to be  bremsstrahlung
from MeV electrons colliding with dense clouds. This mechanism
is usually accompanied by line emission, but no evidence for 
emission line from G26.6$-$0.1 is found.
Assuming a wide band spectrum of G26.6$-$0.1 to be  the same
as $\gamma$-Cygni, we estimate the 0.1--2~GeV band flux to be
6.1$\times 10^{-8}$~photons~cm$^{-2}$s$^{-1}$.
This value is smaller than the EGRET detection limit in our 
survey regions \citep{esposito}.
Therefore, no counterpart of a MeV source in the EGRET third 
catalog \citep{hartman} near at  G26.6$-$0.1 gives no constraint
on the above discussion.
Thus to establish the non-thermal bremsstrahlung origin 
for G26.6$-$0.1,
detection of molecular clouds and MeV-$\gamma$ source
near at G26.6$-$0.1 is essential.

\subsection{How Many SNRs with Hard X-rays in Our Galaxy ?}
\label{discuss3}

We have found 4 diffuse X-ray sources, and 
their hard X-ray spectra are not consistent with SNRs dominated by thermal
emission, but suggest either SN~1006-like, plerionic SNRs, or possibly SNRs
dominated by non-thermal bremsstrahlung.

The {\it ASCA} Galactic plane survey covered the region of
$|l| \leq 45\fdg0$ and $|b| \leq 0\fdg4$, but about 
10\% of the fields should be excluded for the faint source survey
due to the stray lights from bright point sources.
In the above survey region, we found 5 new SNRs;
the 3 sources in this paper, G28.6$-$0.1 by \citet{bamba2001a}, 
and G347.3$-$0.5 by \citet{koyama1997}.
Therefore the number of expected SNRs in the surveyed field is
$5\times\frac{1}{1-0.1}\sim 5.6$.
If we assume that the spatial density of SNR is uniform
in the inner Galactic disk
of $|l| \leq 60\fdg0$ and $|b| \leq 1\fdg0$ field,
then the expected number is $5.6\times\frac{120}{90}\times\frac{2}{0.8} = 19$.
Since the X-ray surface brightness of the new SNR is only 2--3 times of the
background Galactic ridge emission, we may miss detections of more samples 
with lower surface brightness (see \S~\ref{discuss2}). 
Thus the number of SNRs with non-thermal X-rays in our Galaxy
would fur exceed 20.

\acknowledgments

We thank the anonymous referee for helpful comments.
The authors are grateful to the {\it ASCA} Galactic plane survey team.
Our particular thanks are due to
K. Torii, J. Yokogawa, and K. Ebisawa
for their fruitful discussions and comments.
This search made use of the SIMBAD database operated by CDS at Strasbourg,
France.
A.B. and M.U. are supported by JSPS Research Fellowship for Young Scientists.

\appendix
\section{Serendipitous Sources}
\label{appendix}

\begin{figure}[hbtp]
\epsscale{0.32}
\plotone{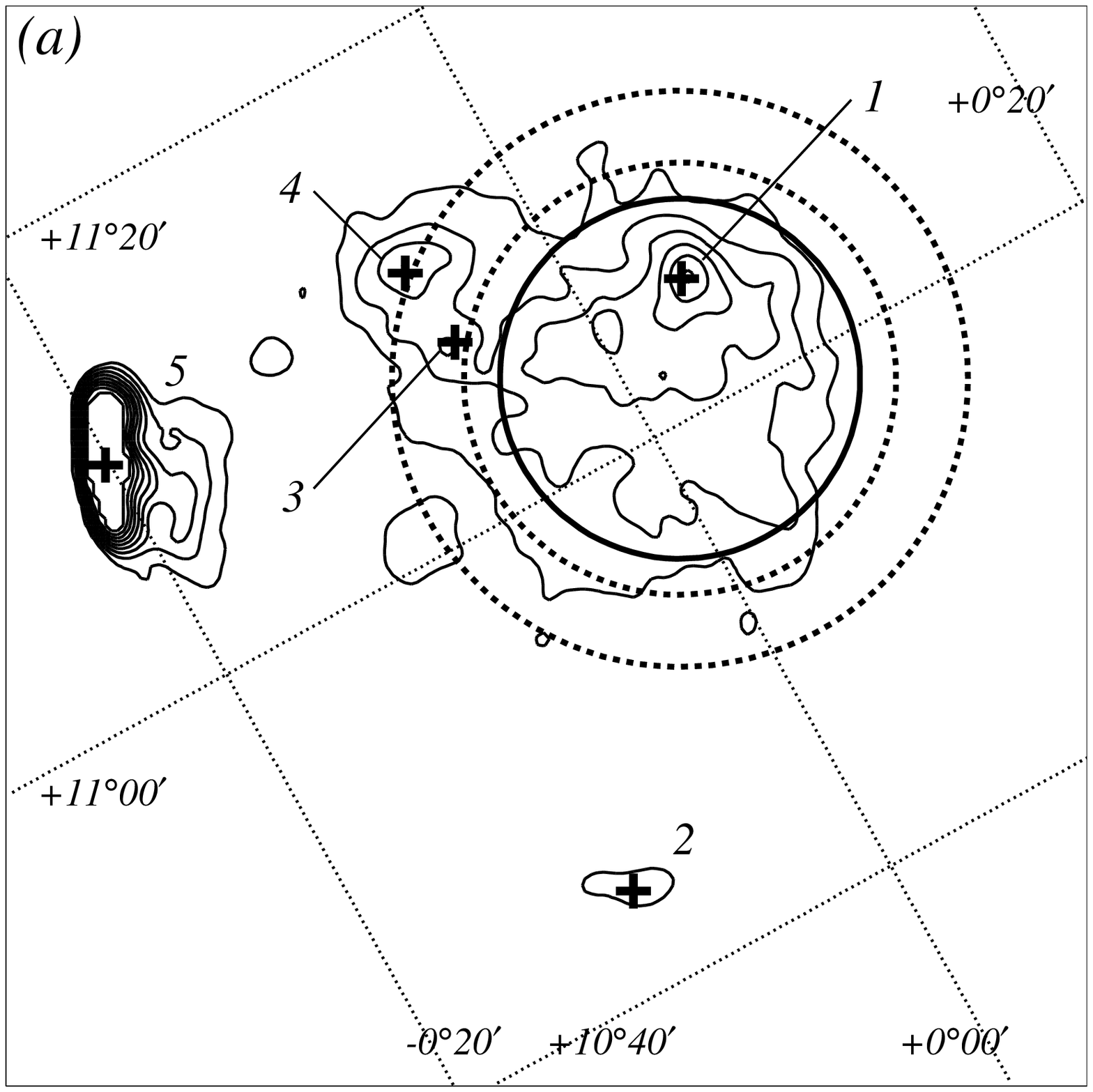}
\plotone{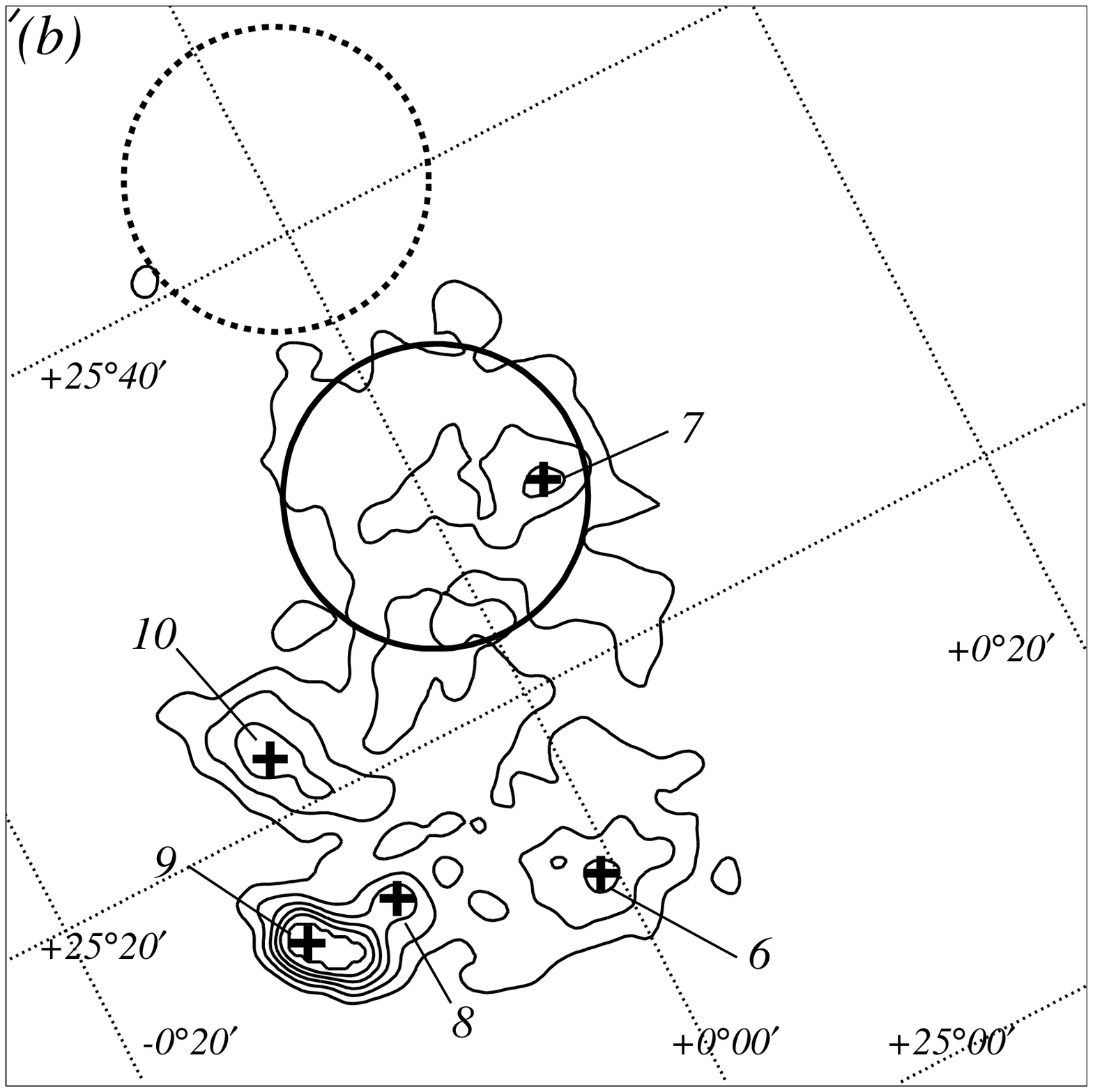}
\plotone{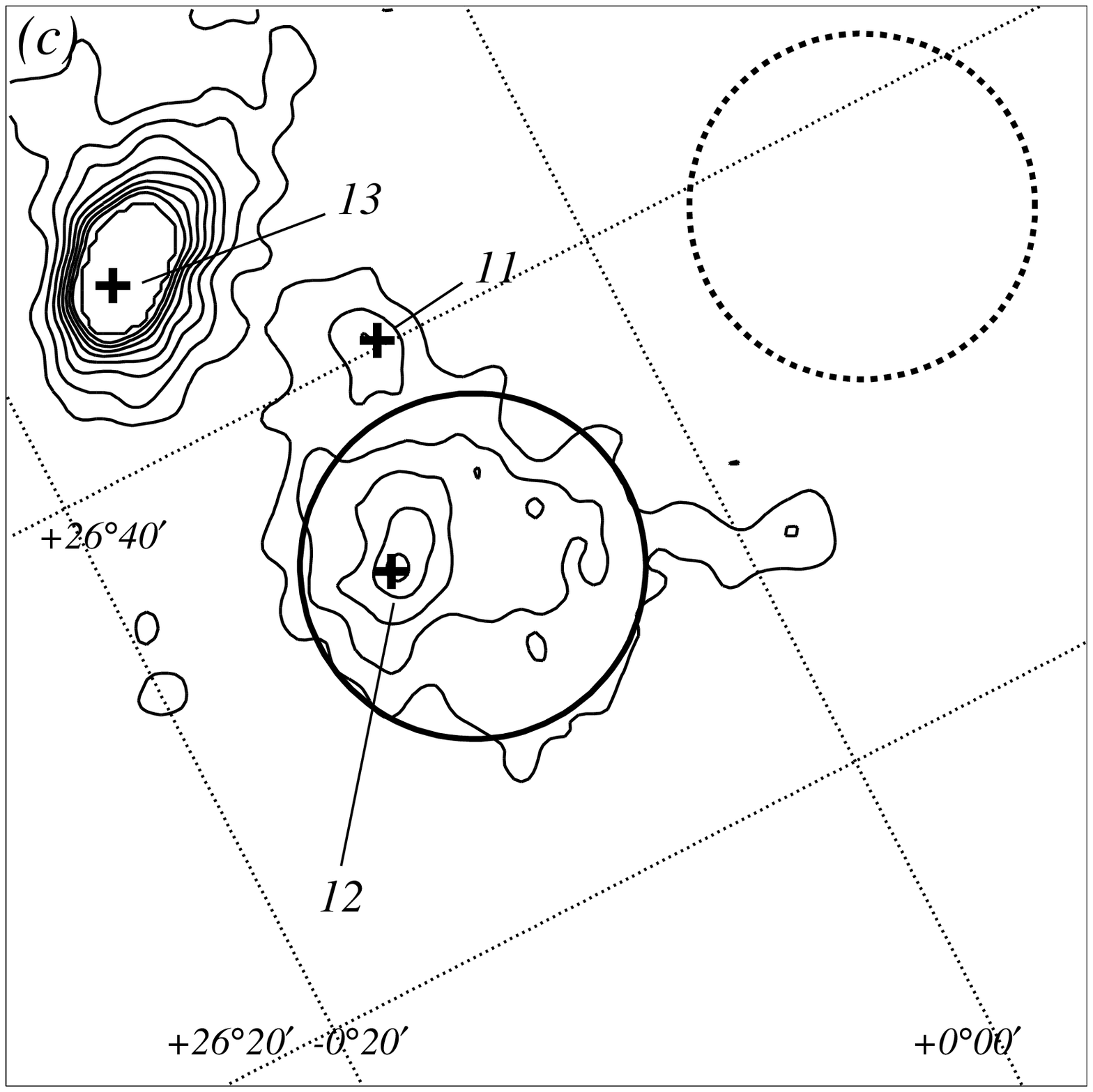}
\caption{The GIS images with Galactic coordinates
around G11.0+0.0 (a), G25.5+0.0 (b), and G26.6$-$0.1 (c)
in the 0.7--7.0~keV band, where exposure time and vignetting effect
are corrected.
The images are smoothed with a Gaussian profiles of $\sigma$ = 0.5~arcmin.
The contour levels are linearly spaced with
6.4$\times 10^{-4}$~counts~arcmin$^{-2}$ interval
from the lowest level of
1.9$\times 10^{-3}$ counts arcmin$^{-2}$s$^{-1}$ for G11.0+0.0,
1.5$\times 10^{-3}$ counts arcmin$^{-2}$s$^{-1}$ for G25.5+0.0,
and 2.4$\times 10^{-3}$ counts arcmin$^{-2}$s$^{-1}$ for G26.6$-$0.1. 
For bright point sources, the higher contour levels
are truncated.
The source and background regions for the spectral study (see text)
are shown with the solid and dashed-line circles.
The crosses indicate the position of point-like sources
shown in Appendix (Table~\ref{other}).
\label{images}}
\end{figure}

\begin{figure}[hbtp]
\epsscale{0.32}
\plotone{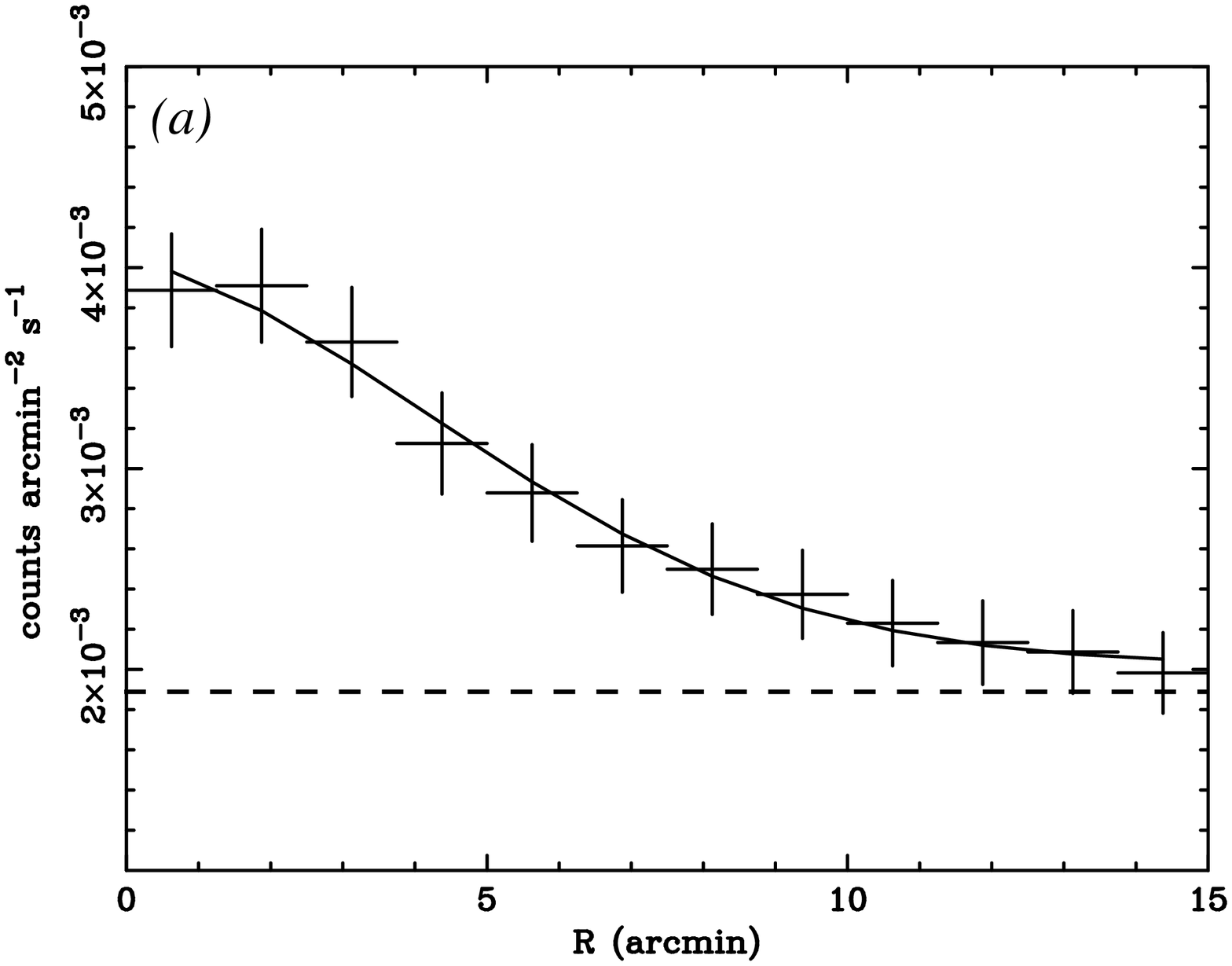}
\plotone{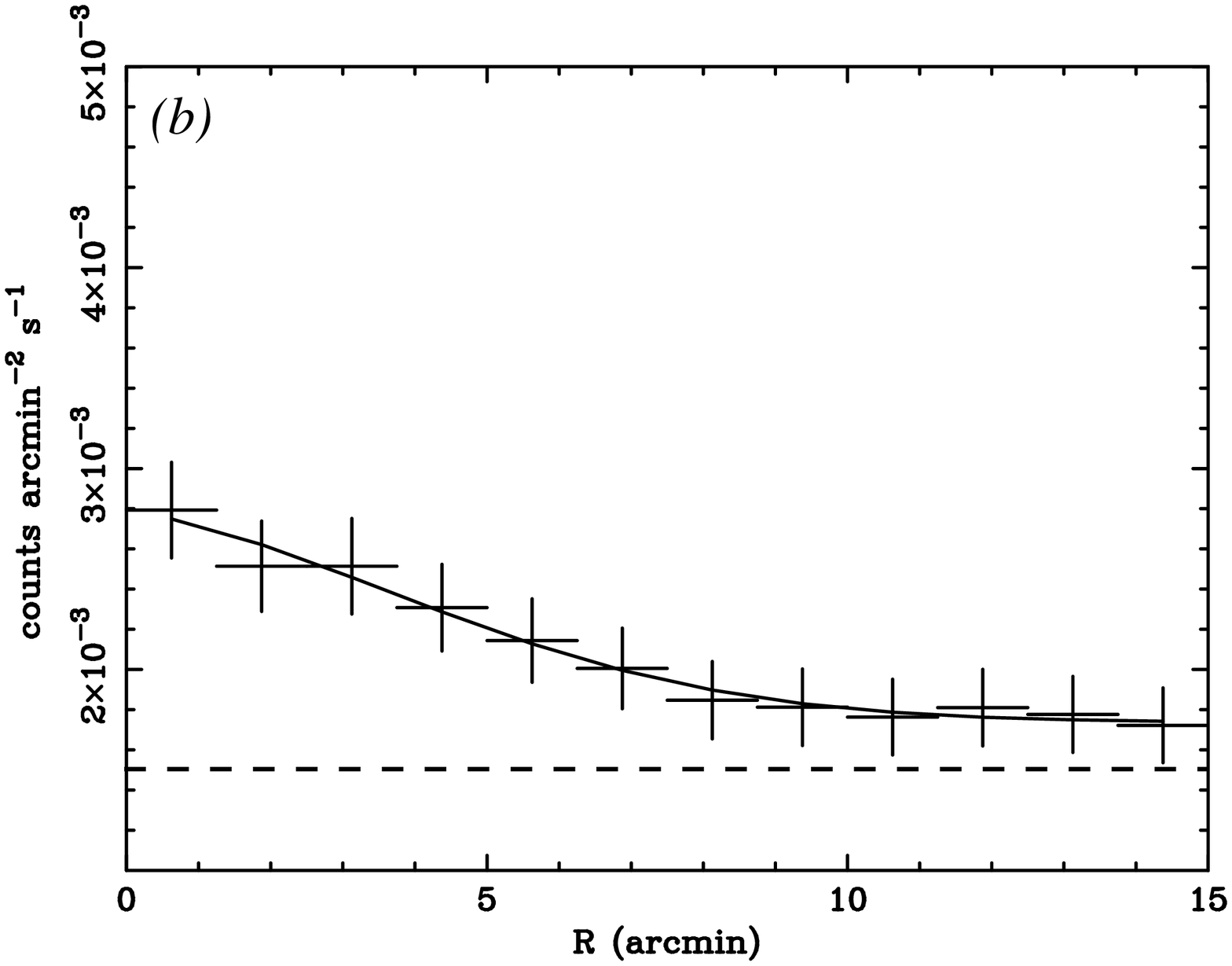}
\plotone{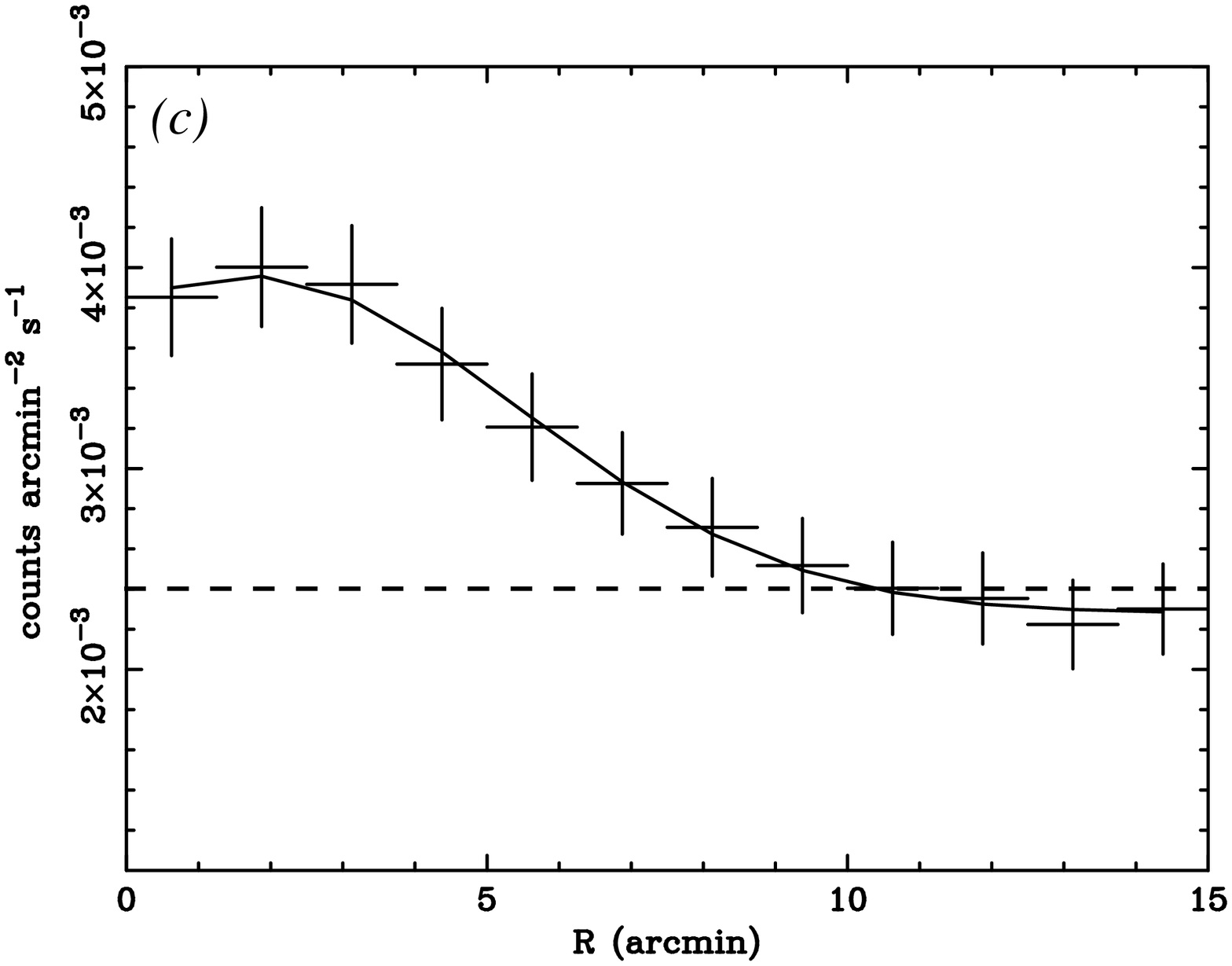}
\caption{The radial profiles (crosses) of
G11.0+0.0 (a), G25.5+0.0 (b), and G26.6$-$0.1 (c)
after correcting the  vignetting effect.
Solid lines are the best-fit curves (see text).
The dashed lines show the lowest contour level 
in Figure~\ref{images}.
The units of horizontal and vertical axes are
arcmin from the center and counts~arcmin$^{-2}$~s$^{-1}$, respectively.
\label{profiles}}
\end{figure}

Thirteen point-like sources above the 6$\sigma$ threshold are found
in the same GIS fields as shown with crosses in Figure~\ref{images}.
Some have been already reported by \citet{sugizaki2001},
while the others are newly reported sources.

We performed the spectral analyses of the point-like sources.
The source regions were selected as 3~arcmin circles
except for sources No.1 and No.8.
For these two sources, 
we made the source regions with 
the 1.5~arcmin circular regions around the sources, because 
the contaminations from G11.0+0.0 are large.
We made  background spectra from  source free regions
near the sources. 
We fitted the background-subtracted spectra with a power-law function and
the results are summarized in Table~\ref{other}.

\onecolumn

\begin{deluxetable}{ccccccc}
\tabletypesize{\scriptsize}
\tablecaption{Log of
Galactic plane survey and follow-up deep observations.
\label{obslog}}
\tablewidth{0pt}
\tablecolumns{2}
\tablehead{
\colhead{Position (J2000)} & \colhead{Date (UT)} & \colhead{Exposure} & \colhead{Obs. Type} & \colhead{Obs. No.}\\
(RA, DEC) & yyyy/mm/dd & [ks]
}
\startdata
(18$^{\rm h}$09$^{\rm m}$50\fs4, $-$19\arcdeg24\arcmin39\farcs6) & 1996/08/04 & 10.3 & survey & 1\\
(17$^{\rm h}$46$^{\rm m}$52\fs8, $-$30\arcdeg02\arcmin31\farcs2) & 1999/09/28 & 39.0 & follow-up & 2\\
(18$^{\rm h}$37$^{\rm m}$48\fs0, $-$06\arcdeg36\arcmin41\farcs8) & 1997/10/14 & 12.4 & survey & 3\\
(18$^{\rm h}$37$^{\rm m}$45\fs6, $-$06\arcdeg36\arcmin43\farcs2) & 1999/09/28 & 37.3 & follow-up & 4\\
(18$^{\rm h}$39$^{\rm m}$43\fs2, $-$05\arcdeg42\arcmin57\farcs6) & 1997/04/20 & 9.8 & survey & 5\\
(18$^{\rm h}$40$^{\rm m}$14\fs4, $-$05\arcdeg40\arcmin44\farcs4) & 1999/10/03 & 35.9 & follow-up & 6\\
\enddata
\end{deluxetable}

\begin{deluxetable}{p{15pc}cccc}
\tabletypesize{\scriptsize}
\tablecaption{The best-fit parameters of the diffuse sources\tablenotemark{a}.
\label{bestfit_para}}
\tablewidth{0pt}
\tablehead{
\colhead{} & \colhead{G11.0+0.0} & \colhead{G25.5+0.0} & \colhead{G26.6$-$0.1} & \colhead{G28.6$-$0.1\tablenotemark{b}}
}
\startdata
Power-law Model\\
\hspace*{5mm}Photon Index\dotfill & 1.6 (1.4--1.9) & 1.8 (1.6--2.2) & 1.3 (1.2--1.5) & 2.1 (1.7--2.4) \\
\hspace*{5mm}Absorption Column Density [$\rm 10^{22}\ H\ cm^{-2}$]\dotfill & 0.8 (0.5--1.1) & 2.4 (1.8--3.2) & 0.4 (0.2--0.6) & 2.7 (2.0--3.5) \\
\hspace*{5mm}Flux\tablenotemark{c}\ [$\rm ergs\ cm^{-2}\ s^{-1}$]\dotfill & $3.8\times 10^{-12}$ & $2.0\times 10^{-12}$ & $3.5\times 10^{-12}$ & $1.8\times 10^{-12}$ \\
\hspace*{5mm}$\chi^2$/degrees of freedom\dotfill & 44.2/42 & 55.4/48 & 73.2/55 & 45.4/47 \\
$srcut$ Model\tablenotemark{d}\\
\hspace*{5mm}Break Frequency [$\times 10^{19}$Hz]\dotfill &  3.2 ($>$ 0.4) & 0.4 (0.07--36) & $1.3\times 10^3$ ($>$ 12) & \nodata \\
\hspace*{5mm}Flux Density\tablenotemark{e} [mJy]\dotfill & 9.0 (8.4--9.6) & 9.6 (8.8--11) & 5.6 (5.2--5.9) & \nodata \\
\hspace*{5mm}Absorption Column Density [$\rm 10^{22}\ H\ cm^{-2}$]\dotfill & 0.8 (0.6--0.9) & 2.3 (2.0--2.7) & 0.6 (0.4--0.7) & \nodata \\
\hspace*{5mm}$\chi^2$/degrees of freedom\dotfill & 45.3/42 & 55.3/48 & 75.6/55 & \nodata \\
NEI Model\\
\hspace*{5mm}Temperature $[{\rm keV}]$\dotfill & 11.5 (7.3--21.0) & 7.1 (4.5--14.0) & 14.5 (8.7--34.8) & 4.4 (3.3--6.2) \\
\hspace*{5mm}Abundance\dotfill & 0.7 (0.1--1.8) & 0.4 (0.1--0.7) & 0.5 (0.1--1.1) & 1.7 (0.5--3.4)\\
\hspace*{5mm}$n_{\rm e}t$ $\rm [10^{9}\ s\ cm^{-3}]$\dotfill & 0.9 ($<$ 4.2) & $3.2\times 10^2$ ($>58$) & 5.1 (3.0--9.1) & 1.1 (0.4--1.9)\\
\hspace*{5mm}Absorption Column Density [$\rm 10^{22}\ H\ cm^{-2}$]\dotfill & 0.6 (0.4--1.1) & 2.3 (1.8--3.0) & 0.9 (0.5--1.4) & 3.9 (3.0--4.9) \\
\hspace*{5mm}Flux\tablenotemark{c}\ [$\rm ergs\ cm^{-2}\ s^{-1}$]\dotfill & $3.8\times 10^{-12}$ & $2.0\times 10^{-12}$ & $3.4\times 10^{-12}$ & $1.7\times 10^{-12}$\\
\hspace*{5mm}$\chi^2$/degrees of freedom\dotfill & 42.7/40 & 48.8/46 & 63.5/53 & 31.1/45\\
\enddata
\tablenotetext{a}{Parentheses indicate single-parameter
90\% confidence regions.}
\tablenotetext{b}{Cited from \citet{bamba2001a}.}
\tablenotetext{c}{In the 0.7--10.0~keV band.}
\tablenotetext{d}{The radio spectral index is fixed to 0.5.}
\tablenotetext{e}{At 1~GHz band.}
\end{deluxetable}

\begin{deluxetable}{p{7pc}cccc}
\tabletypesize{\scriptsize}
\tablecaption{The physical parameters of the diffuse sources.
\label{phys_para}}
\tablewidth{0pt}
\tablehead{
\colhead{} & \colhead{G11.0+0.0} & \colhead{G25.5+0.0} & \colhead{G26.6$-$0.1} & \colhead{G28.6$-$0.1\tablenotemark{a}}
}
\startdata
Distance\tablenotemark{b} [kpc]\dotfill & 2.6 & 7.8 & 1.3 & 7.0 \\
Diameter [pc]\dotfill & 11 & 27 & 4.5 & 20 \\
Luminosity\tablenotemark{c} [ergs s$^{-1}$]\dotfill & $3.7\times 10^{33}$ & $2.3\times 10^{34}$ & $8.1\times 10^{32}$ & $2.2\times 10^{34}$ \\
\enddata
\tablenotetext{a}{Cited from \citet{bamba2001a}.}
\tablenotetext{b}{Calculated from the best-fit absorption column
assuming that  the mean density in the Galactic plane
is 1~H~cm$^{-3}$ }
\tablenotetext{c}{In the 0.7--10.0~keV band.}
\end{deluxetable}

\begin{deluxetable}{p{12pc}ccc}
\tabletypesize{\scriptsize}
\tablecaption{The parameters for thermal plasma scenario.
\label{thermal}}
\tablewidth{0pt}
\tablehead{
\colhead{Parameters} & \colhead{G11.0+0.0} & \colhead{G25.5+0.0} & \colhead{G26.6$-$0.1}
}
\startdata
Emission measure ($E.M.$)\tablenotemark{a}[cm$^{-3}$]\dotfill & $2.3\times 10^{56}$ & $1.5\times 10^{57}$ & $4.9\times 10^{55}$ \\
Dynamical time ($t$)\tablenotemark{b} [s]\dotfill & 1.3$\times 10^{11}$ & 3.9$\times 10^{11}$ & 4.7$\times 10^{10}$ \\
Electron density ($n_{\rm e}$) [cm$^{-3}$]\dotfill & 0.11 & 0.070 & 0.18 \\
Total mass ($M$) [$M_\odot$]\dotfill & 1.8 & 18 & 2.3 \\
Thermal energy ($E$)\tablenotemark{c} [ergs]\dotfill & 1.2$\times 10^{50}$ & 7.3$\times 10^{50}$ & 1.9$\times 10^{49}$ \\
\enddata
\tablenotetext{a}
{$E.M.$ = $n_{\rm e}^2V$, where $n_{\rm e}$ is the electron density
and $V$ is the plasma volume.}
\tablenotetext{b}
{The ratio of the  radius and the sound velocity.}
\tablenotetext{c}
{$E = \frac{3}{2}$($n_{\rm i}$+$n_{\rm e})VkT$,
where $n_{\rm i}$ is the ion density.}
\end{deluxetable}

\begin{deluxetable}{p{1.5pc}ccccc}
\tabletypesize{\scriptsize}
\tablecaption{The best-fit parameters of serendipitously detected sources%
\tablenotemark{a}.
\label{other}}
\tablewidth{0pt}
\tablecolumns{2}
\tablehead{
\colhead{No.} & \colhead{Name} & \colhead{Photon Index} & \colhead{Absorption Column} & \colhead{Flux\tablenotemark{b}} & \colhead{Reduced $\chi^2$} \\
 & & & [$\rm 10^{22}\ H\ cm^{-2}$] & [$\rm ergs\ cm^{-2}\ s^{-1}$]
}
\startdata
1\dotfill & AX~J1809.7$-$1918 & 3.1(1.9--5.0) & 1.6 (0.4--3.3) & 3.1$\times 10^{-13}$ & 9.1/16 \\
2\dotfill & AX~J1809.8$-$1943 & 8.8 ($>$5.9) & 1.7 (0.7--2.5) & 6.1$\times 10^{-13}$ & 9.8/20  \\
3\dotfill & AX~J1810.4$-$1921 & 2.2 (1.5--3.2) & 0.6 (0.1--1.4) & 6.3$\times 10^{-13}$ & 15.4/17 \\
4\dotfill & AX~J1810.5$-$1917 & 3.4 (2.4--4.8) & 0.5 ($<$1.3) & 5.7$\times 10^{-13}$ & 11.7/17 \\
5\dotfill & AX~J1811.4$-$1926 & \nodata & \nodata & \nodata & \nodata \\
6\dotfill & AX~J1837.3$-$0652 & 2.3 (1.6--3.2) & 5.9 (3.8--9.0) & 1.8$\times 10^{-12}$ & 43.1/42  \\
7\dotfill & AX~J1837.4$-$0637 & 0.8 (0.5--1.1) & 0.5 (0.1--1.2) & 9.3$\times 10^{-13}$ & 48.8/59  \\
8\dotfill & AX~J1837.5$-$0653 & 1.1 (0.5--1.9) & 5.3 (2.8--8.8) & 4.1$\times 10^{-12}$ & 20.6/29  \\
9\dotfill & AX~J1838.0$-$0655 & 0.8 (0.4--1.2) & 4.0 (2.8--5.7) & 1.1$\times 10^{-11}$ & 12.9/12  \\
10\dotfill & AX~J1838.1$-$0648 & 2.1 (1.6--2.8) & 3.9 (2.7--5.6) & 1.5$\times 10^{-12}$ & 48.3/47 \\
11\dotfill & AX~J1840.4$-$0537 & 3.4 ($>$1.6) & 0.6 ($<$4.0) & 1.4$\times 10^{-13}$ & 6.8/13 \\  
12\dotfill & AX~J1840.4$-$0545 & 2.3 (0.9--4.7) & 5.8 (1.7--17.9) & 6.2$\times 10^{-13}$ & 6.6/13 \\
13\dotfill & AX~J1841.0$-$0536 & 1.0 (0.3--1.9) & 3.2 (1.1--6.4) & $2.1\times 10^{-11}$ & 13.2/21 \\
13\dotfill & flare (obs.6) & 1.1 (0.9--1.4) & 7.2 (6.0--8.6) & 9.5$\times 10^{-11}$ & 43.2/50 \\
\enddata\
\tablenotetext{a}{Parentheses indicate single-parameter
90\% confidence regions.}
\tablenotetext{b}{In the 0.7--10.0~keV band.}
\tablecomments{
No.1: \citet{sugizaki2001} identified this source
to an A type star HD166077, although the X-ray spectrum was
much harder than those of normal stars. In this
paper, we selected the background region around the source
to remove the hard X-ray contamination from G11.0+0.0
and obtained a softer spectrum, consistent with a normal star.\\
No.2: The count rates and photon indices
with {\it ROSAT} \citep[1RXS~J189951.5194345]{voges} and {\it ASCA} \citep{sugizaki2001}
are consistent with the present results.\\
No.4: AX~J1810.5$-$1917 was brighter than AX~J1810.4$-$1921 in obs.1,
but was fainter in obs.2 (see Table~\ref{obslog}), hence 
is a transient or a variable source.\\
No.5: This source is  a bright SNR both in the radio and X-ray bands \citep{dowens}
with a pulsar at the center \citep{torii}.
No pulsation is found with the FFT and epoch-folding search,
which may be due to the limited statistics.
Also no spectral analyses is available
due to location at the edge of the GIS FOV.\\
No.9: This is the  $Einstein$, IPC \citep{hertz} and {\it ASCA} \citep{sugizaki2001} source.
The count rates in all the observations have been constant.\\
No.13: This is a transient X-ray pulsar detected
in both obs.5 (quiescent) and 6 (quiescent+flare).
The $4.7~s$-pulsation is found at the flare phase
\citep{bamba1999,bamba2001b}.}
\end{deluxetable}

\end{document}